\documentclass{aa} 
\usepackage{graphics,latexsym,amssymb,times,psfig} 
 
\def\gsim{ \lower .75ex \hbox{$\sim$} \llap{\raise .27ex \hbox{$>$}} }  
\def\lsim{ \lower .75ex\hbox{$\sim$} \llap{\raise .27ex \hbox{$<$}} }  
 
\begin{document} 
 
\title{Aborted jets and the X--ray emission of radio--quiet AGNs} 
 
 
\author{Gabriele Ghisellini \inst{1}, Francesco Haardt \inst{2}, 
and Giorgio Matt \inst{3}  
} 
 
\offprints{G. Ghisellini; gabriele@merate.mi.astro.it} 
\institute{ 
Osservatorio Astronomico di Brera, via Bianchi 46, I--23807 Merate, Italy; 
\and Dipartimento di Scienze di Como, Universit\`a dell'Insubria, 
via Valleggio 11, Como, I--22100  Italy; 
\and Dipartimento di Fisica, Universit\`a degli Studi Roma Tre, Via della Vasca 
Navale 84, I--00146 Roma, Italy. 
} 
 
\date{Received ---} 
 
\titlerunning{Aborted jets in AGNs} 
\authorrunning{G. Ghisellini, F. Haardt \& G. Matt}

\abstract{  
We propose that radio--quiet quasars and Seyfert galaxies have 
central black holes powering outflows and jets which propagate 
only for a short distance, because the velocity of the ejected 
material is smaller than the escape velocity. 
We call them ``aborted" jets. 
If the central engine works intermittently, blobs of material may be 
produced, which can reach a maximum radial distance and then fall back, 
colliding with the blobs produced later and still moving outwards.  
These collisions dissipate the bulk kinetic energy of the blobs by   
heating the plasma, and can be responsible (entirely or at least in part)  
for the generation of the high energy emission in radio--quiet objects.  
This is alternative to the more conventional scenario in which   
the X--ray spectrum of radio--quiet sources originates in a hot  
(and possibly patchy) corona above the accretion disk.  
In the latter case the ultimate source of energy of the emission of both   
the disk and the corona is accretion.  
Here we instead propose that the high energy emission is powered   
also by the extraction of the rotational energy of the black hole   
(and possibly of the disk).  
By means of Montecarlo simulations we calculate
the time dependent spectra and light curves,
and discuss their relevance to the X--ray spectra
in radio--quiet AGNs and galactic black hole sources.
In particular, we show that time variability and spectra are 
similar to those observed in Narrow Line Seyfert 1 galaxies.
\keywords{
Accretion disks --- 
Radiation mechanism: thermal --- 
X--rays: galaxies --- 
Galaxies: jets ---
Galaxies: Seyfert }  
}  
\maketitle

\section{Introduction}  

One of the most popular scenario to explain the dichotomy between   
radio--loud and radio--quiet Active Galactic Nuclei (AGNs) assumes   
that only rapidly spinning black holes can  give rise 
to the relativistic jets responsible for the  radio 
emission and higher frequency non--thermal radiation observed in  
radio--loud objects (e.g. Blandford 1990). 
Therefore it was with some surprise that the first evidence 
for rapid rotation of a black hole came from a radio--quiet  
object, namely MGC--6--30--15.  
In the X--ray spectrum of this Seyfert 1 a broad iron line  
was observed by ASCA and {\it Beppo}SAX (Tanaka et al. 1995; 
Guainazzi et al.  1999) to be consistent with emission outside 
6$R_{\rm g}$ ($R_{\rm g}=GM/c^2$), i.e. the innermost stable orbit 
of an accretion disc around a static black hole.   
Recent XMM--Newton observations (Wilms et al. 2001; Fabian et al. 2002),   
however, indicate that the emission may extend well within 6$R_{\rm g}$,  
thus requiring a rotating black hole (the innermost stable orbit in the  
case of a  maximally rotating black hole being $R_{\rm g}$), and confirming the  
result obtained by Iwasawa et al. (1996) during a low flux state of the  
source observed with ASCA.  
Moreover, the very steep radial dependence of the iron line emissivity was   
interpreted by Wilms et al. (2001) as evidence for the extraction of the spin 
energy of  a Kerr black hole, even if a pure geometrical explanation  
(but still requiring a rotating black hole) is possible (Martocchia et al. 2002),  
provided that the illuminating source is very close to the black hole and resides  
on the symmetry axis.  
Similar evidence comes from XMM--Newton observations of the   
galactic black hole candidate XTE J1650--500 (Miller et al. 2002a).  
  
Furthermore, Elvis, Risaliti \& Zamorani (2002) have recently suggested that   
the X--ray background requires a high efficiency of mass to energy  
conversion in the accretion process, possible if the black hole  
is rotating, but problematic in the case of a Schwarzschild black hole.  
  
These results are at odds with the idea of a slowly spinning black hole in   
radio--quiet objects.   
On the other hand, and in a complementary way, radio--quiet 
objects are not radio--silent: even if the  
``dichotomy" between radio--loud and radio--quiet objects is currently 
under scrutiny (see e.g. White et al. 2000 and Ivezic et al. 2002), 
all AGNs can produce radio emission at some level, 
which in turn is consistent with the idea that some sort of jet  
or outflow is always present, responsible to accelerate electrons to   
relativistic energy to radiate by the synchrotron process in the radio band.  
This idea has received recently fully support by  
VLBI imaging of Seyfert galaxies (Ulvestad 2003, and references therein) 
which revealed the presence of a mini--jet (at the sub--pc scale) 
in many Seyfert (radio--quiet) galaxies.  
In several cases it was also possible to detect the proper motions 
of knots in the jet, which appear to move with subluminal apparent 
velocities of the order of a tenth of the speed of light. 
 
Therefore it is conceivable to assume that all black hole plus 
accretion disk systems in AGNs can produce some kind of  
outflow or jet, but that only in a minority of cases (i.e. the  
``pure" radio--loud objects) the jet is successfully launched and  
accelerated to relativistic speeds.  
In the majority of cases, the jet is ``aborted", yet it is   
responsible for a relatively weak radio emission.  
The idea that all AGNs produce a jet is not new: among others, 
Falcke \& Biermann (1995) suggested the jet--disk symbiosis 
for all AGNs, while Henri \& Petrucci (1997) and Malzac et al. (1998)  
have argued  that the initial part of a jet in a radio--quiet object 
can produce relativistic particles illuminating the disk.  
In these scenarios, however, the jet has either bulk relativistic motion 
or it contains very energetic particles, which are relativistic in the 
jet--comoving frame. 
In our scenario, as it will be explained below, the jet has 
sub--relativistic 
bulk velocities by assumption, and also most of the emitting 
electrons are thermal, with subrelativistic temperatures. 
 
It is possible that a source which is usually radio--quiet  
may occasionally be successful in launching relativistic jets.  
This could explain the properties of galactic superluminal sources,  
in which major outflows sometimes occur.  
If this is true, these sources should be considered   
a crucial link between radio--loud and radio--quiet objects. 
In this respect, it is worth noting that in Galactic superluminal 
sources and Galactic black hole candidates there is often (even 
in radio--quiet states) the presence of a high energy X--ray 
power law, which may be associated with the emission from a jet, 
or at least from an outflow (as in the case of XTE 1118+480: 
Miller et al. 2002b).   
  
The aim of this paper is twofold.  
First we will explore if the simplest ``abortion mechanism"  
which comes into mind, i.e. a ``jet" which does not  
succeed to reach the escape velocity, can work, at least qualitatively.  
Then we explore the possibility that the power initially in the   
jet and/or outflow can be used to heat the particles responsible  
to emit the X--ray flux from radio--quiet AGNs.  
In other words, we substitute the popular hot corona, possibly patchy,  
which sandwiches the accretion disk (i.e. Haardt \& Maraschi 1991),  
with a single hot region on the rotation axis of the black hole,  
thought to be the site of the jet abortion. 
We then perform numerical simulations assuming to launch
many blobs with slightly different velocities and time separations,
calculate their trajectories and follow their evolution, accounting
for the collisions occurring between them.
This allow us to calculate the produced luminosity in each
collision, and the total luminosity received by the observer
in the likely case that more than one shell--shell collision
is occurring at any given observing time.
We will also study if the typical scattering optical depths 
and temperatures of the scattering particles in the aborted jet 
scenario are in agreement with what observed (Petrucci et al. 2001; 
Perola et al. 2002).  
Finally we will discuss our findings and derive some observational  
consequences enabling to test this scenario.

\section{Escape velocity from a Kerr black hole}  
  
The mechanism to form, accelerate and collimate jets in radio--loud   
objects is not well understood, even if several proposals exist   
in the literature (see e.g. the review by Lynden--Bell 2001).  
Here we do not attempt to propose a new mechanism. 
We simply postulate that in radio loud--quiet objects a similar  
mechanism is at work, but it is, on average, not able to impart  
to the outflowing material a bulk velocity larger than the  
escape velocity.   
We consider a blob of material in ballistic motion, and neglect
the influence of magnetic fields and/or radiative fields
(for calculations including the effect of accretion disk radiation
see Vokrouhlick\`y \& Karas 1991).

\begin{figure}  
\vskip -0.4 true cm  
\psfig{file=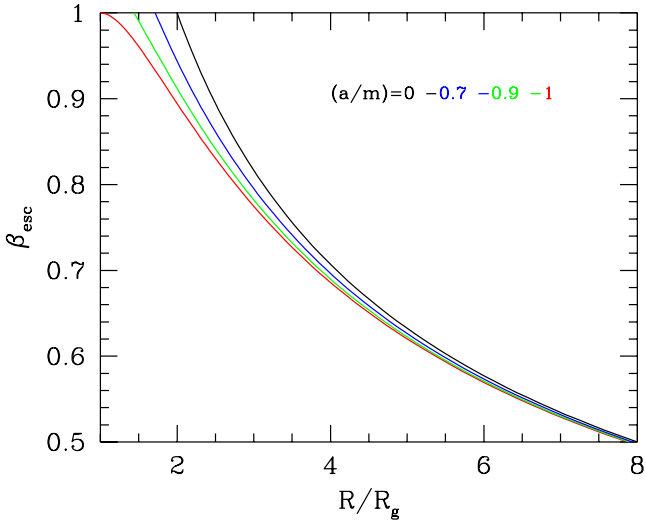,width=.52\textwidth}  
\vskip -2 true cm  
\caption{Escape velocity as a function of distance form the black hole  
and for $a=$0, 0.7, 0.9, and 1, from top to bottom.} 
\label{fig_besc} 
\end{figure}

\begin{figure}  
\psfig{file=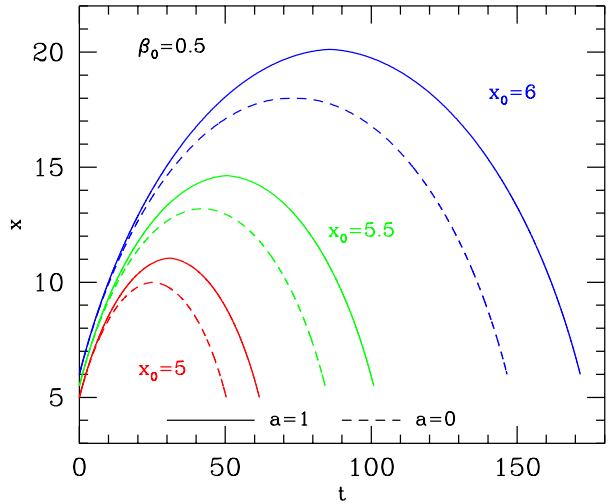,width=.53\textwidth}  
\vskip -2 true cm  
\caption{Trajectories of test particles in motion along the rotation  
axis of an extreme Kerr hole ($a=1$, solid lines), and Schwarzschild   
(dashed lines), for the initial radii $x_0\equiv R/R_{\rm g}$=5, 5.5,   
and 6, as labeled.  
The initial velocity is $\beta_0=0.5$.  
Time is measured in units of the initial light crossing time  
$R_0/c$.  
}  
\label{fig_xt}
\end{figure}

The equation of motion of a test particle along the rotation axis 
of a Kerr hole is given by (see e.g. Vokrouhlick\`y \& Karas 1991):  
\begin{equation}  
{d\beta\over dx}\, =\, - \, {x^2 -a^2 \over \beta\gamma^2  
(x^2+a^2-2x)(x^2+a^2)}   
\label{dbdx}
\end{equation}  
where $x\equiv R/R_{\rm g}$, $a$ is the (dimensionless)  
specific angular momentum ($\sim$1 for a maximally rotating Kerr hole),  
and $R_{\rm g}=GM/c^2$ is the gravitational radius.  
Its solution is given by:  
\begin{equation}  
\gamma(x)\, = \, \gamma_0\, \left[  
{(x^2 + a^2)\over (x_0^2+a^2)} \, {(x_0^2-2x_0 +a^2)   
\over (x^2 -2x+a^2)}\right]^{1/2}  
\label{gammax}
\end{equation}  
It is then possible to find the escape velocity  
for a test particle having an initial velocity  
$c \beta_0$ at the distance $x_0\equiv R_0/R_{\rm g}$ on the   
rotational axis of a Kerr hole:  
\begin{equation}  
\beta_{\rm esc} \, =\, \left[{ 2 x_0 \over x_0^2 +a^2}\right]^{1/2}
\label{besc}  
\end{equation}  
yielding $\beta_{\rm esc}=1$ for $R=R_{\rm g}$ and $a=1$  
(maximal Kerr hole),  
and for $R=2R_{\rm g}$ and $a=0$ (Schwarzschild hole).  
Fig. \ref{fig_besc} shows the escape velocity $\beta_{\rm esc}$ as a   
function of the initial radius $x_0$ for different values of   
the angular momentum parameter $a$.  
Material with $\beta<\beta_{\rm esc}$ will then outflow in the radial  
direction reaching a maximum distance from the black hole, and  
then it will fall back.  
The maximum distance above the hole where the motion inverts, $x_{\rm max}$,  
can be found setting $\gamma(x_{\rm max})=1$ in Eq. (\ref{gammax}), to find  
\begin{eqnarray}  
x_{\rm max} \, &=&\, { k+\sqrt{k^2(1-a^2)+2ak-a^2} \over k-1}; \nonumber \\  
k\, &\equiv&\, { x_0^2+a^2 \over \gamma_0^2(x_0^2+a^2-2x_0)}  
\label{xmax}
\end{eqnarray}  
Fig. \ref{fig_xt} shows for illustration some examples of ``trajectories".  

\section{Jet power and internal shocks}  

In this section we discuss the basic features of our model.
In order to have an analytical treatment, we will
consider here only a simple case, while the more
realistic, but much more complex case, will be 
treated numerically in section 5.

The total power carried by the jet is  
\begin{equation}  
L_{\rm j}\,  =\, \Gamma_0 \dot M_{\rm out} c^2  
\label{Lj}
\end{equation}  
where $\Gamma_0$ is the bulk Lorentz factor   
corresponding to the initial velocity of the jet,    
and $\dot M_{\rm out}$ is the mass ejection rate,  
which is in principle a free parameter of the model,  
but it can be constrained by the observed ratio  
between the X--ray and UV luminosity.  
A fraction $\eta_{\rm j}$ of $L_{\rm j}$ will be   
transformed into radiation.  
  
We propose that the central engine producing the jet is not   
working continuously, but intermittently, leading to collisions   
between blobs of material launched at different times, and   
moving with slightly different velocities.  
In the case of successful jets this will lead to   
the formation of  ``internal shocks", 
first proposed by Rees (1978) for the jet of the radio--galaxy M87, 
later proposed to explain the prompt emission of gamma ray bursts  
(Rees \& M\'esz\'aros 1992; Rees \& M\'esz\'aros 1994; Sari \& Piran 1997),  
and recently proposed again to explain the entire spectral energy 
distribution of all radio--loud AGNs (Ghisellini 1999; Spada et al. 2001). 
In this scenario the shells collide at a typical distance $R_{\rm diss}$  
from the black hole given by $R_{\rm diss}\sim \Gamma^2 R_0$,  
where $\Gamma$ is the average bulk Lorentz factor of the shells and $R_0$  
is the initial shell--shell separation.  
In the case of aborted jets the intermittency of the process will lead  
to collisions between a blob already falling backwards  
to the hole and a blob still moving upwards.  
There is no need, in this case, to assume slightly different velocities,  
since the collisions will occur anyway.  
The details of the collision depend upon the initial velocities and the  
typical ``launching" site, which is related to the time needed to   
accelerate a blob.  
  
There may be a range of locations $R_{\rm coll}$ where collisions   
preferentially take place, depending on the initial time separation   
of two consecutive shells, their initial velocity   
$\beta_0$, and on the initial launching site $R_0$.  
These parameters, together with the angular momentum  
of the Kerr hole, determine $R_{\rm coll}$, which in turn determines  
the amount of bulk kinetic energy which can be dissipated.  
Larger $R_{\rm coll}$ in fact correspond to smaller kinetic  
energies (and larger potential ones).  
Fig. \ref{fig_scontro} shows for illustration simple examples of the velocity  
and kinetic energy of two colliding shells, as a function   
of $R_{\rm coll}$.   
In these examples we assume the same mass, initial   
velocity and launching site for all blobs.
  
In a more realistic case, $R_{\rm coll}$ will be dependent partly on   
the exact values of the initial parameters, and partly on the past history   
of the process, in the sense that a falling blob may interact with more   
than one later blobs (including some which may have $\beta>\beta_{\rm esc}$).  
Interestingly, the power which can be extracted is a function of  
the dissipation distance, since the more powerful collisions will be  
between just launched blobs and falling blobs with a back velocity  
of the order of the initial one, and this will occur closest to the hole  
(see Fig. \ref{fig_scontro}).  
We also stress that in these examples we have only treated the simplest  
possible case, neglecting in particular the interaction of the blob  
with the radiation produced by the accretion disk and the motion  
off the angular momentum axis.

With our simple approximations, the dissipated power is the entire  
kinetic power of the shells when they collide, since they  
have equal and oppositely directed velocities.   
Thus the total momentum is zero if the shells have, as we assume, equal masses.  
We can then link the radiative efficiency of   
the process to the kinetics of the collision through:  
\begin{equation}  
\eta_{\rm j}\, =\, \epsilon_{\rm e}\,   
{\Gamma_{\rm coll} -1 \over \Gamma_0 -1} 
\label{etaj} 
\end{equation}  
where $\epsilon_{\rm e}$ is the fraction of the collisional energy  
radiated by the electrons (and possibly electron--positron pairs),  
and $\Gamma_{\rm coll}$ is the bulk Lorentz factor when the  
blobs collide.  
We do not know what is the dominant acceleration mechanism  
transforming ordered into random energy, and this   
precludes to know with confidence what fraction of the total   
energy goes to the emitting electrons (and possibly positrons),  
and what fraction is instead given to protons and to magnetic field.   
Equipartition between protons, electrons and magnetic field  
would result in a value of $\epsilon_{\rm e}\simeq 1/3$.  
The value of $\eta_{\rm j}$ can be calculated for any assumed value  
of $\epsilon_{\rm e}$ from Fig. \ref{fig_scontro} and Eq. (\ref{etaj}). 
 
Observationally, the X--ray luminosity is of the order of 10--50  
per cent of the luminosity in the optical--UV (e.g. Walter \&  
Fink 1993), thought to be produced by the accretion disk, for  
which $L_{\rm d} =\eta_{\rm d} \dot M_{\rm in} c^2$.  
Here $\dot M_{\rm in}$ is the mass accretion rate and  
$\eta_{\rm d}$ is the accretion efficiency.  
 
If all the X--ray luminosity $L_X$ comes from the aborted jet,   
i.e. $L_X=\eta_{\rm j} L_{\rm j}$, while $L_{\rm d} \sim L_{\rm UV}$,  
we have:  
\begin{equation}  
{\dot M_{\rm out} \over \dot M_{\rm in}} \, =\,   
{ 1 \over \Gamma_0}\, {\eta_{\rm d} \over \eta_{\rm j}} \,  
{ L_X \over L_{\rm UV} }  
\label{dotmout}
\end{equation}  
Since $\Gamma_0$ is always of order unity, we have that  
$\dot M_{\rm out}\sim (L_X/L_{\rm UV})\dot M_{\rm in} \sim 0.1 \dot M_{\rm in}$  
if the jet and the accretion efficiencies are of the same order.  
  
  
  
\section{The X-ray emission}

In this section we again discuss the very simplified case
of the collision of two oppositely directed blobs
having the same mass and same velocity (in modulus),
in order to understand the basics of the interaction
with an analytical treatment.

\subsection{Formation of the X--ray continuum}  
  
Having established that the simple aborted jet scenario   
discussed above can in principle account for the energetics  
of the X--ray emission, we must now check whether the physical  
parameters of the aborted jets, i.e. its optical depth and  
temperature, are consistent with the X--ray spectral constraints.  
  
The accelerated leptons are embedded in a very large radiation energy density  
produced by the close--by accretion disk and by the local magnetic field 
possibly enhanced by the collision. 
Under these conditions, the most efficient radiation processes are  
Inverse Compton scattering and cyclo--synchrotron emission.  
We will see below that energy balance ensures that the equilibrium lepton  
energy is mildly relativistic at most.  
This implies that the cyclo--synchrotron process occurs in the   
self--absorbed regime and then does not contribute to the cooling.  
This process could however be important for establishing  
an electron Maxwellian distribution in a timescale   
shorter than the dynamical time (Ghisellini, Haardt \& Svensson, 1998).  
For simplicity, we then assume that the energized leptons have a  
thermal energy distribution.  It is worth noting
that a distribution which is not a perfect Maxwellian,  
but has a well defined mean energy, such as a distribution  
with a well defined peak energy, or  a power law energy  
distribution ($\propto E^{-n}$) with $n>3$, give rise to  
a Comptonization spectrum which cannot be distinguished by  
the one formed by a perfect Maxwellian   
(see e.g. Ghisellini, Haardt \& Fabian 1993).

To find the typical average energy $\gamma m_{\rm e}c^2$ of the emitting 
leptons, we balance radiation losses with the   
energy gains due to the shell--shell collision process.  
The inverse Compton cooling rate is:  
\begin{equation}  
\dot E_{\rm c} \, =\, {4\over 3} \sigma_{\rm T} c U_{\rm r} \gamma^2\beta^2  
\label{dotec}
\end{equation}  
where $U_{\rm r}$ is the radiation energy density of the disk emission. 
The heating rate due to the shell--shell collision is   
\begin{equation}  
\dot E_{\rm h}\, =\, {\eta_{\rm j} L_{\rm j} \over N_{\rm e}}
\label{doteh}  
\end{equation}  
where $N_{\rm e}$ is the total number of leptons in the emitting region.  
Balancing heating and cooling we have:  
\begin{equation}  
\gamma^2\beta^2 \, =\,  {3\over 4}\, {\eta_{\rm j} L_{\rm j} \over  
\sigma_{\rm T} c N_{\rm e} U_{\rm r}}  
\label{g2b2}
\end{equation}  
Assume that the ``jet" emitting region has a transverse size $r_{\rm j}$  
and a width $\Delta r_{\rm j}$.  
Its scattering optical depth is   
\begin{equation}  
\tau_{\rm T} \, =\, {N_{\rm e}\over f r_{\rm j}^2\Delta r_{\rm j}}\,  
\sigma_{\rm T} \Delta r_{\rm j}   
\label{tau}
\end{equation}  
where $f$ is a geometry dependent factor (e.g., $f=4\pi/3$ for a sphere).

\begin{figure}  
\psfig{file=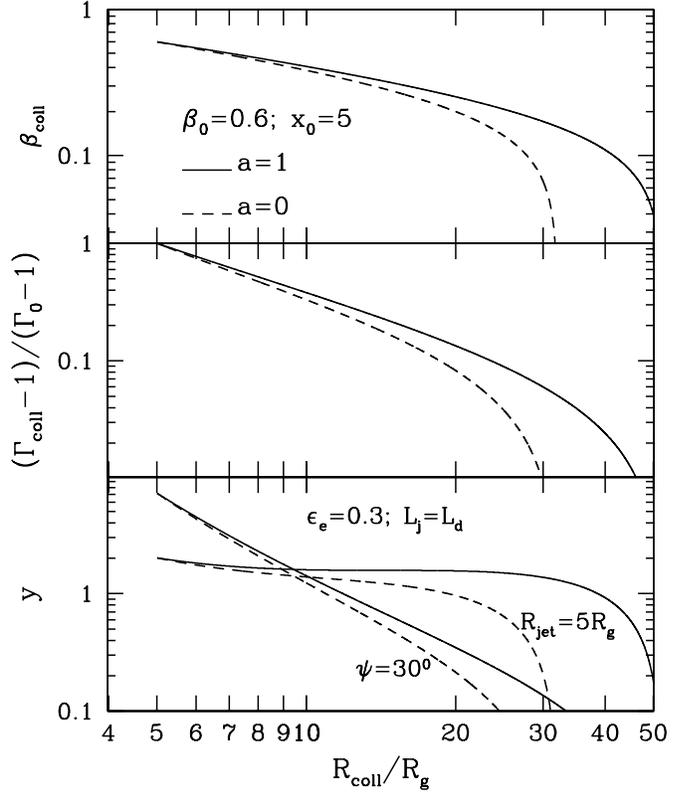,width=.63\textwidth}  
\vskip -0.5 true cm  
\caption{ 
Consecutive shells collide at different distances from the black hole  
and with different velocities.  
In the collision they dissipate their kinetic energy,  
proportional to $\Gamma_{\rm coll}-1$.  
An initial distance $x_0=5$ and initial velocity $\beta_0=0.6$  
is assumed.   
Solid lines refer to a maximally rotating Kerr hole,  
dashed line to a Schwarzschild black hole.  
Top panel shows the velocity of the shells when they collide;  
the mid panel shows the total dissipated energy, which is greater   
closer ot the hole;  
the bottom panel shows the Comptonization $y$ parameter, calculated   
following Eq. (\ref{y}), and assuming, as labeled, $L_{\rm j}=L_{\rm d}$  
and $\epsilon_{\rm e}=0.3$.   
We show the case of shells of fixed dimensions (set to $5R_{\rm g}$ in 
this example) or expanding as $r_{\rm j}=\psi R_{\rm coll}$, 
where we have used $\psi=30^\circ$. 
}  
\label{fig_scontro}
\end{figure}  
If $\tau_{\rm T} \lsim 1$, the Comptonization parameter is  
$y\equiv (4/3)\tau_{\rm T} \gamma^2\beta^2$.  
In this case we have:  
\begin{equation}  
y \, =\, {\eta_{\rm j} L_{\rm j} \over f r_{\rm j}^2c U_{\rm r} }  
\label{y}
\end{equation}  
The energy density $U_{\rm r}$ depends on the specific geometry   
of the system and on the accretion luminosity.  
Note that $L_{\rm d}$ should always be greater than $\eta_{\rm j} L_{\rm j}/2$,  
since about half of the jet produced luminosity impinges on the accretion disk.  
However, observations typically give $L_{\rm d}\sim L_{UV}\sim 10 L_X$,   
suggesting that the accretion--produced luminosity is dominating  
with respect to the jet luminosity intercepted and reprocessed by the disk.  
For simplicity, let us assume a Newtonian disk producing a total 
power $L_{\rm d}$, which has a minimum radius $r_*$. 
Per unit surface area the dissipated flux is (e.g., Frank, King \& Raine 1985): 
\begin{equation} 
D(r)\, =\, { 3 L_{\rm d} r_* \over 4 \pi r^3 } 
\, \left[ 1- \left( {r_*\over r}\right)^{1/2}\right], 
\label{dr}
\end{equation} 
where $r$ is the radial coordinate of the accretion disk. 
The intensity $I(r)$ corresponding to this dissipation is 
$I(r) = D(r) /\pi$. 
At the location $R_{\rm coll}$ the total radiation energy density 
produced by the disk is 
\begin{eqnarray} 
U_{\rm r}(R_{\rm coll})  &=&{2\pi \over c} \int_0^{\mu_{\rm max}} I(r) d\mu 
\nonumber \\ 
&=& 
{ 3 L_{\rm d} r_* R_{\rm coll} \over 2 \pi  c } \, \int^\infty_{r_*}  
{  1- \left( {r_*/ r}\right)^{1/2}  \over 
r^2 \left( R^2_{\rm coll} + r^2 \right)^{3/2} } dr 
\label{ur}
\end{eqnarray} 
where $\mu\equiv \cos\theta$ and $\theta$ is the angle with respect to the 
normal of the accretion disk. 
 
%
%
 
Inserting $U_{\rm r}(R_{\rm coll})$ in Eq. (\ref{y}) we can calculate the  
Comptonization $y$ parameter. 
In the bottom panel of Fig. \ref{fig_scontro} we show, as illustration, how 
$y$ changes as a function of $R_{\rm coll}$ if the shell remains of the  
same dimension (i.e. $r_{\rm j}$= constant) or if instead it expands 
as a cone of semiaperture angle $\psi$ (in this case we have 
assumed $r_{\rm j} =\psi R_{\rm coll}$). 
In both cases we show the behavior for a maximally rotating Kerr hole 
($a\sim 1$) and for a Schwarzchild hole ($a=0$). 
 
A constant $r_{\rm j}$ implies an almost constant 
$y$--parameter. 
This is because $\eta_{\rm j}$ and $U_{\rm r}$  
decrease with $R_{\rm coll}$ in approximately the same way. 
If the shells expand, instead, $y$ decreases because  
the optical depth of the scattering electrons decreases as  
$\tau_{\rm T} \propto r_{\rm j}^{-2}\propto R_{\rm coll}^{-2}$.

Eq. (\ref{y}) also gives the $y$--parameter in the classical  
case of an homogeneous corona sandwiching the accretion disk.  
Assuming that the jet dissipates the entire available gravitational  
energy, and that half of it is reprocessed by the accretion disk,
we can substitute $\eta_{\rm j}L_{\rm j}$ with $L_X$,  
$L_{\rm d}$ with $L_X/2$ 
Furthermore, the disk radiation energy density in this case 
can be approximated by  
$U_{\rm r} = 2 L_{\rm X} /(\pi r_{\rm j}^2 c)$. 
Setting $f=\pi/2$, we finally obtain $y=1$.

%

\subsection{Are electron--positron pairs important?}  
  
The assumption that the jet carries a power $L_{\rm j}$   
in the form of kinetic energy links the amount  
of transported power with the scattering optical depth.  
On the other hand, $\tau_{\rm T}$ may not be linearly  
proportional to $L_{\rm j}$, because of the possible presence of  
electron positron pairs, which contribute fully to the scattering   
optical depth but not so much to the jet power (if protons are also   
present and are dominating the jet inertia).  
We will therefore estimate first a lower limit to the   
scattering optical depth assuming no pairs, and then  
we will make some considerations to evaluate the   
contribution of pairs.


To calculate the initial optical depth of each blob,
without the contribution of pairs, we first consider the
kinetic energy carried by each blob, and use it to find its mass $M_i$:
\begin{equation}
M_i\, =\, {L_{\rm j} \Delta t \over \Gamma_{i,0} c^2}
\label{mass}
\end{equation}
where $\Delta t$ is the time interval between two consecutive
blob ejection.
All the values of the quantities in Eq. \ref{mass}
have to be considered as average values.
The total number of electrons contained in each blob is
therefore $N_{\rm e} = M_i/m_{\rm p}$ if the jet is made
by an electron--proton plasma.
Inserting this value in Eq. \ref{tau} we get:
\begin{equation}
\tau_{\rm T}\, =\, {\sigma_{\rm T} L_{\rm j} \Delta t \over m_{\rm p} c^2
f r_{\rm j}^2 \Gamma_{i,0} c^2}\, =\,
{4 \pi \over f\Gamma_{\rm i, 0}} \,
{L_{\rm j}\over L_{\rm Edd}} \,
\left( {r_{\rm g} \over r_{\rm j}}\right)^2 \,
{ \Delta t \over r_{\rm g}/c}
\label{tau2}
\end{equation}
For the second equality, we have used the jet luminosity in units
of the Eddington one, the blob size in units of $r_{\rm g}$ and
time in units of $r_{\rm g}/c$.
In this way it becomes clear that the optical depth
(without the contribution of pairs) is scale invariant with respect
to the mass of the black hole.
Furthermore, it is also clear that, for $r_{\rm j}$ of the order
of the gravitational radius, for jet kinetic powers of order
of 0.01--0.1 of the Eddington luminosity and for time intervals of a few
times $r_{\rm g}/c$ the initial optical depth is of order of 0.01--0.1.

If the jet, instead, is made by a pure electron--positron plasma, 
then the initial value of the optical depth is a factor $m_{\rm p}/m_{\rm e}$
greater, and consequently $\tau_{\rm T}$ becomes much larger than unity.
In this case the pairs annihilate efficiently, since the annihilation timescale
is of order of $r_{\rm j}/(c\tau_{\rm T}$), and most of the kinetic
power of the jet is lost through annihilation.
We conclude that if the jet is energetically important,  
then its inertia must be given mainly by protons.

We stress that there is a remarkable feature in the model,  
which is the link between the kinetic power and the amount of  
electrons and protons carried by the jet.  
In pair--corona models, the optical depth of the  
corona has a lower limit given by the created pairs 
(Haardt \& Maraschi 1991).  
Here, instead, the limit is given by the jet power.  

We can also estimate the possible importance of pairs
created during the emission phase.
In other words, when primary
electrons are heated by the shell--shell collision,
they can emit photons above the pair production threshold:
if the source is sufficiently compact, these photons
create pairs which increase the optical depth and
contribute to the emission.

The effects of pairs have been studied in detail assuming  
a steady source in pair equilibrium (creation equal annihilation, 
Svensson 1984).  
Our source is probably never in steady state, but we can use the  
results of these studies as a guide to estimate the relevant  
pair optical depths and temperatures.  
The optical depth due to pairs, produced in steady state and in pair 
equilibrium without pair escape, can be approximated by (Haardt 1994):  
\begin{equation}  
\tau_{\pm} \, \simeq\, 0.016 \, (\ell_{\rm x})^{0.56};\qquad   
20\lsim \ell_{\rm x}\lsim 200  
\label{taupairs}
\end{equation}  
where $\ell_{\rm x} \equiv \sigma_{\rm T} L_{\rm x}/ (r_{\rm j} m_{\rm e} c^3)$ 
is the compactness of the X--ray radiation produced 
by a shell--shell collision, and can be considered equal to
$\eta_{\rm j} \ell_{\rm j}$.
This equation assumes that the radiation process is Comptonization,
and is valid assuming spectral indices $\alpha_x$ of the 
Comptonized spectrum around unity. 
The range of X--ray compactnesses relevant for Seyfert galaxies  
is within the same range of the compactness relevant  
for the accretion disk emission, i.e. $\ell_{\rm x}\sim$1--10. 

By comparing Eq. \ref{taupairs} with Eq. \ref{tau2} we see
that pairs should be unimportant for compact jets.
Furthermore, as will be clear in the following section, the
optical depth of each shell is increasing each time
it collides with another one (if lateral expansion can be negleted),
driving the typical optical depth to values a factor $\sim$10
larger than the initial value.
We then conclude that electron positron pairs
do not play a fundamental role in this scenario.

\section{Numerical simulations}

We have discussed so far the illustrative and very simple case 
of a pair of blobs of equal mass, equal launching site and initial
velocity, colliding at different distances from the black hole.
In a more realistic case, we should assume that the time interval 
$\Delta t$ between the launch of consecutive blobs is variable,
as well as the blob mass, initial velocity, and their launching site.
To simulate this, we use a Monte Carlo code which extracts the 
initial quantities within an assigned distribution.
We have then assumed a Gaussian distribution
for the initial velocity 
(i.e. we assign a mean value and a width $\sigma$), while for the
time interval $\Delta t$ we assume a Poisson distribution.
We have then considered the case of a Gaussian distribution of the
blob masses, leading to initially different kinetic energies 
of the blobs, and the case of equal kinetic energies for all blobs.
In this latter case the mass is therefore determined by the 
blob velocity.

We follow the trajectory of all blobs, and when they collide we
calculate the corresponding dissipation through the equation of 
conservation of energy and momentum:
\begin{equation}
\Gamma_1 m_1+\Gamma_2m_2 \, =\, 
\Gamma_f \left( m_1+m_2+\epsilon^\prime/c^2 \right)
\end{equation}
\label{energy}
\begin{equation}
\Gamma_1\beta_1 m_1+\Gamma_2\beta_2 m_2 \, =\,  
\Gamma_f\beta_f \left( m_1+m_2+\epsilon^\prime/c^2 \right)
\label{momentum}
\end{equation}
In these equations the subscripts 1 and 2 stand for the two blobs, 
$\Gamma=(1-\beta^2_f)^{-1/2}$ is the bulk Lorentz factor just before the collision
and $\Gamma_f$ if the final Lorentz factor of the two blobs just after
the collision. $\epsilon^\prime$ is the dissipated energy, calculated
in the frame comoving with the merged blobs. 
Hereinafter primed quantities are calculated in this frame.
A completely anelastic collision is assumed.
The two unknowns are $\beta_f$ (or $\Gamma_f$) 
and $\epsilon^\prime$ which are given by:
\begin{equation}
\beta_f \, =\, {\Gamma_1\beta_1 m_1 + \Gamma_2\beta_2 m_2 \over 
 \Gamma_1m_1 + \Gamma_2m_2}
\label{betaf}
\end{equation}
\begin{equation}
{\epsilon^\prime\over c^2}  = 
\left[ m_1^2 + m_2^2 + 2m_1m_2\Gamma_1\Gamma_2(1-\beta_1\beta_2) \right]^{1/2} 
- m_1 - m_2\, 
\label{epsilon}
\end{equation}
A fraction $\epsilon_{\rm e}$ of $\epsilon^\prime$ is given to the
radiating electrons, making the X--ray luminosity, as observed 
in the comoving frame:
\begin{equation}
L^\prime_x \, \simeq\, \epsilon_{\rm e} \, { \epsilon^\prime  \over 
t^\prime_{\rm diss}} 
\label{lx}
\end{equation}
where $t^\prime_{\rm diss}$ is the time needed to radiate the
energy $ \epsilon_{\rm e} \epsilon^\prime$.
In the dense seed photon environments we are considering,
the electron radiative cooling time is fast, always shorter
than blob--blob crossing time $t^\prime_{\rm cross}$.
We then assume that $t^\prime_{\rm diss}\simeq t^\prime_{\rm cross}$.
We compute $t^\prime_{\rm cross}$ in the frame
of one of the two blobs,
\begin{equation}
t^\prime_{\rm diss} \, \simeq\,  t^\prime_{\rm cross} \, =\,
{\Delta r_{\rm j}\over c} \, { 1-\beta_1\beta_2 \over |\beta_1-\beta_2| }
\label{tdiss}
\end{equation}

We identify $L^\prime_x$ with the ``radiative jet luminosity"
$\eta_{\rm j}L_{\rm j}$ of Eq. (\ref{y}) and derive the corresponding
value of the Comptonization parameter $y$.

At any time, the observer can see more than one collision,
and therefore we calculate, at any time, the number of
colliding shells. 
We then sum up their luminosities
and assume, for each collision, a triangular luminosity
profile with simmetric rise and decay timescales 
(each of duration $t_{\rm diss}$),
whose time integral is equal to the energy dissipated 
by the electrons.
The observed luminosity produced by each collision is modified by
special relativistic effects, which we account for
introducing the Doppler beaming parameter:
\begin{equation}
\delta \, \equiv \, {1 \over \Gamma_f (1-\beta_f\cos\theta)}
\end{equation}
where $\theta$ is the viewing angle and $\beta_f$ must
be considered with its sign (positive for shells approaching 
the observer).
We then multiply the comoving luminosity by $\delta^4$
and divide the intrinsic timescale by $\delta$.
Note that $\delta$ can be greater or smaller than unity,
depending on $\beta_f$ and $\theta$.
We take into account time dilation due to general relativity,
by dividing intrinsic times by the factor 
$(1-2 R_{\rm g} /R)^{1/2}$, 
where $R$ is the distance from the black hole,
and multiplying intrinsic frequency by the same factor.
We call the resulting observed luminosity $L^{\rm obs}_x$.
\begin{figure}
\vskip -0.7 true cm    
\psfig{file=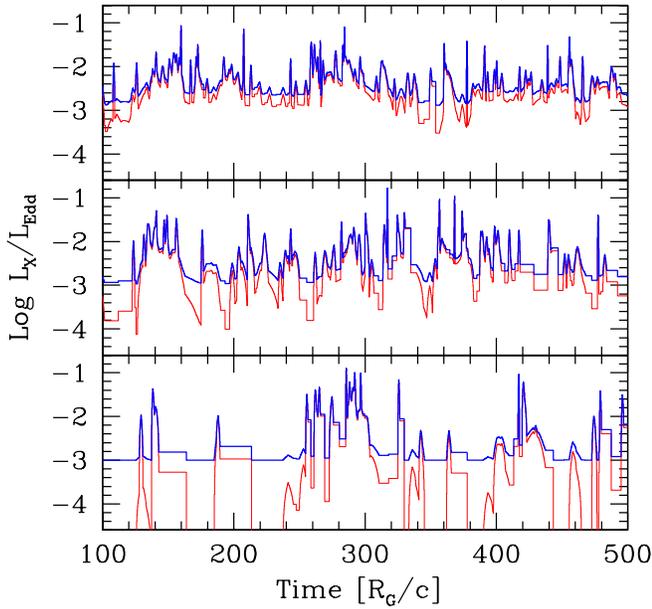,width=.53\textwidth}  
\vskip -0.5 true cm  
\caption{Light curves of the integrated X--ray luminosity
(in units of the Eddington one)
corresponding to numerical simulations, as discussed
in the text. Thin lines corresponds to the power emitted
by shell--shell collisions. To this, we have added a constant
luminosity (0.1\% Eddington) and the sum is shown by the
thick solid lines.
Note that these light curves take into account the total
number of shells which an observe sees ``on" at a given time.
The three cases differ only for the average time interval between
the blobs, which is 4, 8 and 16 $R_{\rm G}/c$ from top to bottom.
Note that these results are scale--invariant: the light curve
does not change changing the black hole mass.
} 
\label{lc_alt}
\end{figure}  

The luminosity is assumed to be produced when two shells
merge. 
For simplicity, we assume that the entire produced radiation
is emitted once the velocity of the merged shell has 
already reached the final value, and use this velocity 
to calculate the appropriate special relativistic effects.
We then continue to follow the merged shells until
a new collision occurs, or until it reaches a distance
from the black hole equal or less than the initial one.

Note that the mass of a generic blob increases for
each collision, and in the absence of side expansion
this implies a corresponding increase of the
scattering optical depth.

We have also taken into account the additional electron heating
due to Coulomb collisions between protons and electrons,
occurring on the timescale $t^\prime_{\rm ep}$ given by
(e.g. Stepney 1983)
\begin{equation}
t^\prime_{\rm ep}\, \sim \, 
{\Delta r_{\rm j}\over c \tau_{\rm T}}\, 
{m_{\rm p}/ m_{\rm e} \over \ln\Lambda }
\left( {kT_{\rm e} \over m_{\rm e}c^2} 
+{kT_{\rm p} \over m_{\rm p}c^2} \right)^{3/2} 
\label{tep}
\end{equation}
where $T_{\rm e}$ and $T_{\rm p}$ are the electron and
proton temperatures, and $\ln\Lambda\sim 20$ is the Coulomb logarithm.
Coulomb collisions between proton and electrons allows
electrons to be heated for a time $t_{\rm ep}$. 
The additional energy gained by the electrons is radiated 
on a cooling timescale. 
This corresponds to an additional luminosity
\begin{equation}
L^\prime_{x,p} \, \simeq\,  {\epsilon^\prime\, \epsilon_{\rm p} \over 
t^\prime_{\rm ep}} 
\label{lxp}
\end{equation}
where $\epsilon_{\rm p}$ is the fraction of dissipated energy 
heating the protons.
We then transform $L^\prime_{x,p}$ as before, taking into account 
beaming and gravitational redshift.
In our cases $t^\prime_{\rm cross}$ is almost always shorter than 
$t^\prime_{\rm ep}$, and consequently $L^\prime_{x,p}$ is almost always smaller than
$L^\prime_x$, but it lasts longer, somewhat smoothing the
lightcurve.

\begin{figure}  
\vskip -0.7 true cm
\psfig{file=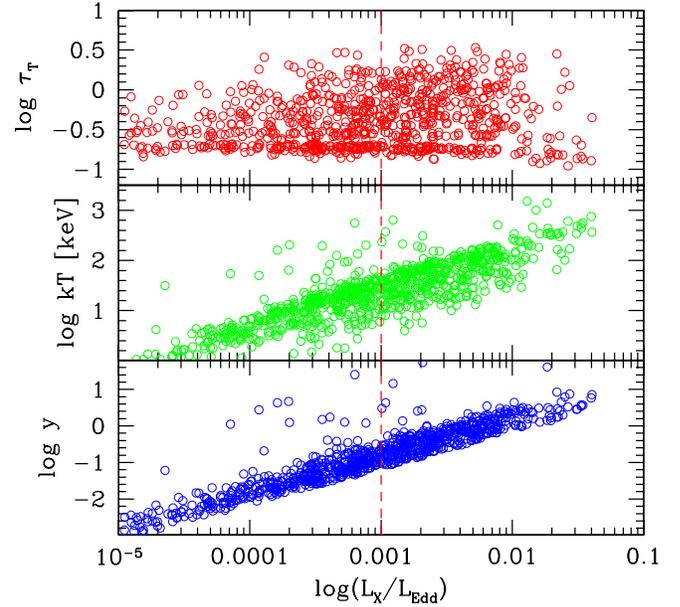,width=.53\textwidth}  
\vskip -0.7 true cm  
\caption{Thomson scattering optical depth, plasma temperature
(in keV) and Comptonization $y$ parameter as a function of integrated
X--ray luminosity produced in single collisions for
the case corresponding to the top panel of Fig. \ref{lc_alt}.
The dashed vertical line corresponds to 0.1\% of the Eddington 
luminosity which we have assumed to be constant. 
}
\label{parameter}
\end{figure}  
\begin{figure}  
\psfig{file=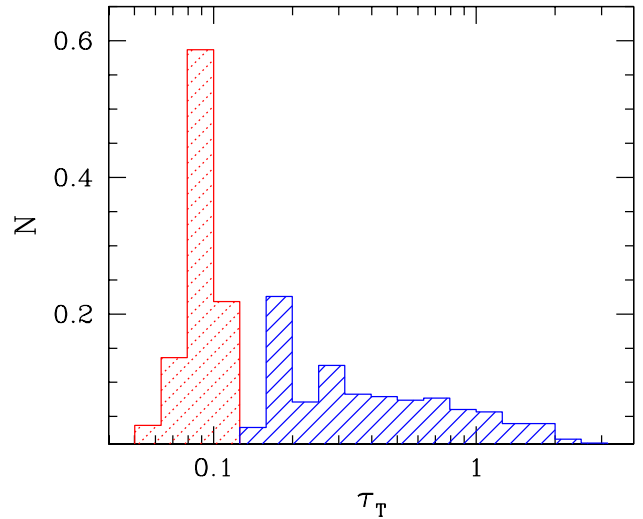,width=.53\textwidth}  
\vskip -1.8 true cm  
\caption{
When a blob is launched, it has an optical
depth belonging to the distribution shown by the dotted
histogram.
The other histogram shows the distribution of $\tau_{\rm T}$
soon after each collision.
}
\label{taufig}
\end{figure}  

For the simulations shown in Fig. \ref{lc_alt} 
and in Fig. \ref{parameter} we have assumed:

\begin{itemize}


\item $\langle \beta_0 \rangle = 0.5 $; $\sigma_{\beta_0} =0.2$; 
Gaussian distribution;

\item $\langle \Delta t \rangle =$4, 8, 16 $R_{\rm g}/c$; Poisson distribution;

\item launching site $R_0=5R_{\rm g}$ equal for all blobs;

\item blob size $r_{\rm j} =4 R_{\rm g}$ equal for all blobs;

\item blob width $\Delta r_{\rm j} =r_{\rm j}$ equal for all blobs;

\item the blobs do not expand nor contract;

\item the jet initial kinetic luminosity $L_{\rm j}$ equal to the
radiative accretion disk luminosity, both being equal to 0.1 $L_{\rm Edd}$;

\item maximally spinning black hole;

\item $\epsilon_{\rm e}=1/3$; $\epsilon_{\rm p}=1/3$ for all blobs;

\item $\cos\theta=0.5$.

\end{itemize}

Clearly, some of these assumptions are not fully justified physically,
but have been done just for ease of computation.
In particular, one expects that the blob size and width change
during the blob trajectory; one furthermore expects that the
launching size is not the same for all blobs.

\begin{figure}  
\vskip -0.7 true cm
\psfig{file=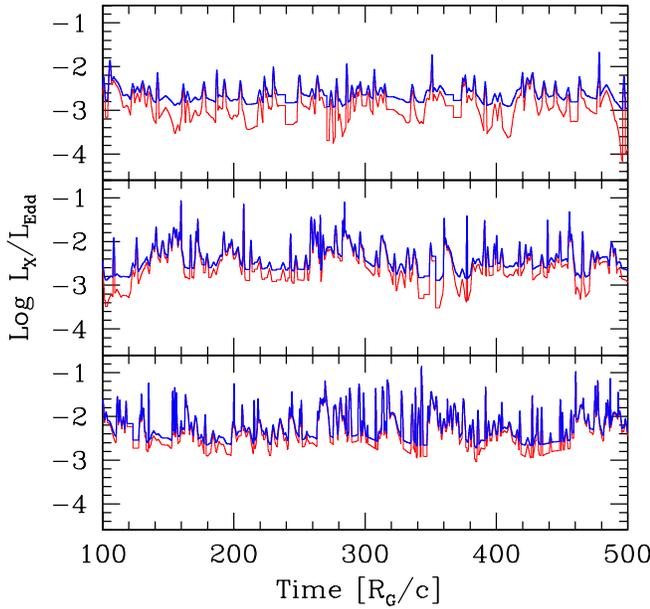,width=.53\textwidth}  
\vskip -0.7 true cm  
\caption{
Light curves of the integrated X--ray luminosity
(in units of the Eddington one)
corresponding to numerical simulations.
The three cases differ for the average initial average
velocity, which is 0.3, 0.5 and 0.7 $c$, from top to bottom.
The average time interval between consecutive shell ejection
is $\langle \Delta t \rangle=4 R_{\rm G}$.
All other parameters are the same as in Fig. \ref{lc_alt}.
}
\label{lc_b0}
\end{figure}  

Fig. \ref{lc_alt} shows the light curves of the bolometric luminosity
for three cases, corresponding to $\langle \Delta t \rangle =$
4, 8, 16 $R_{\rm g}/c$ (from top to bottom).
Being an intrinsically intermittent process, the resulting
light curves (thin lines) show a very large range of variability
(the luminosity can virtually vanish for some time interval).
This corresponds to the case of a completely ``on and off" 
mechanism, i.e. there is a vanishing particle density
between the launched blobs.
We consider this unphysical, even if useful for estimating
the radiated power in the way we have discussed above.
A more sophisticated treatment should take into account
a ``smoother" behavior of the central engine, with density
and velocity profiles described by smooth functions.
While this is referred to future work, the general effect
on the light curve will be to allow some of the energy carried by 
the blobs (that in a more physical scenario can be thought as 
overdense regions) to be dissipated all along the jet, 
and not only during collisions with other overdense region.
This emission may correspond to a minor fraction
of the luminosity produced during each collision, but
it should be much more continuous.

To partly account for that we have added, in Fig. \ref{lc_alt}
and in Fig. \ref{lc_b0},
a constant luminosity equal to 0.1 per cent of the Eddington value.
The thick solid lines are the sum of this contribution and the
contribution produced by the shell--shell collisions.

The three different curves in Fig. \ref{lc_alt} show the effect 
to change the time interval between the launching of blobs.
For increasing time intervals, less spikes per unit time are produced 
and the overall process becomes less efficient, since
there are less collisions.

Note that, in agreement with our simple estimates,
the average emitted luminosity (at least for short $\Delta t$) 
is around a few per cent of the jet initial kinetic power.

Fig. \ref{parameter} shows the optical depth $\tau_{\rm T}$,
the electron temperature $kT$ and the Comptonization $y$ 
parameter as a function of the luminosity produced in each collision
(i.e. this is not the sum over the ``on" shells used in
Fig. \ref{lc_alt}).
For reference, we have drawn a vertical line corresponding to 
0.1 per cent of the Eddington luminosity and corresponding
to the stationary component assumed above.
For this case we have assumed $\Delta t=4 R_{\rm g}/c$ as in the 
top panel of Fig. \ref{lc_alt}.
As can be seen, when the luminosity is relevant (i.e. $L^{\rm obs}_x\sim 10^{-1}
L_{\rm j}=10^{-2} L_{\rm Edd}$), 
we have optical depths between 0.5 and 2, 
temperatures around 100 keV and $y\sim$1.
For this case we obtain values of the optical depth which are
a few times larger than the initial ones, due to
the previous collision done by the shell, as shown in Fig. \ref{taufig}.


\subsection{Observed spectra}

Since we can calculate  the optical depth and
temperature for each shell--shell collision, 
we can calculate the emitted spectrum
assuming thermal Comptonization as the radiative process.
To this end we use the analytical formulae of Titarchuk \& Mastichiadis
(1994), and assume that the soft radiation field is a blackbody
peaking at some energy $\epsilon_0$. 
For all cases discussed in the following, we have assumed $\epsilon_0=5$ eV.
For each observing time, we sum up the spectra of the ``on" shells,
including the contribution of the steady component, whose bolometric
luminosity amounts to the 0.1 per cent of the Eddington luminosity.
We have assumed that the spectral shape of this steady component is
$F(\nu)\propto \nu^{-1} e^{-\nu/\nu_c}$, with $h\nu_c=150$ keV.

This specific chosen spectral index may account for the contribution 
of the hot corona, above the accretion disk, while it is not clear 
if it can be directly associated to the jet generation process 
(e.g., to a particle injection in the jet smoother 
than the assumed on--off mechanism).
In this latter case the usual feedback process operating in the
hot corona model, fixing the spectral index close to unity,
does not work.
In fact, when the power of the steady component is much smaller than the
luminosity of the disk, there is no feedback, and the Comptonization
spectrum should be much steeper than unity.

\begin{figure}  
\vskip -0.7 true cm
\psfig{file=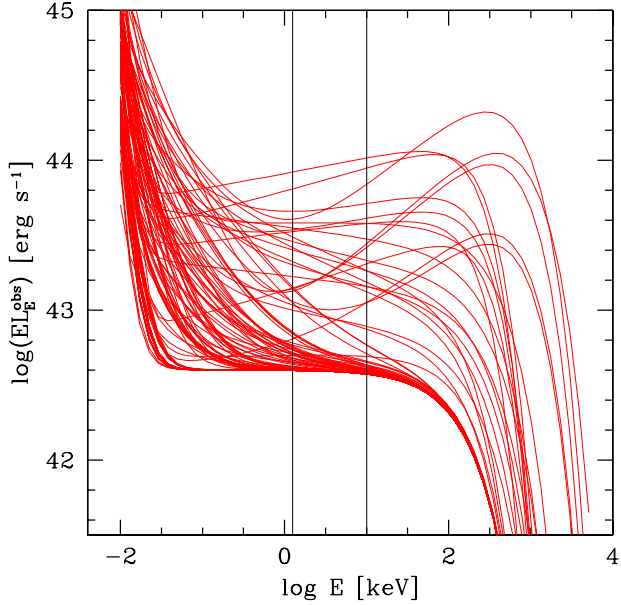,width=.53\textwidth}  
\vskip -0.7 true cm  
\caption{
Example of time dependent spectra for the
case illustrated in the top panel of Fig. \ref{lc_alt},
assuming a black hole of $10^8$ solar masses.
Spectra are calculated every $2.7\times 10^{-2} R_{\rm g}/c$,
but for clarity the figure shows spectra
every $0.53 R_{\rm g}/c$ (i.e. every 800 seconds for 
a black hole of $10^8$ solar masses), from 
$t/(R_{\rm g}/c)=240$ and 300.
Vertical line mark the 1--10 keV band.
}
\label{spectra_r05}
\end{figure}  

\begin{figure}  
\vskip -0.7 true cm
\psfig{file=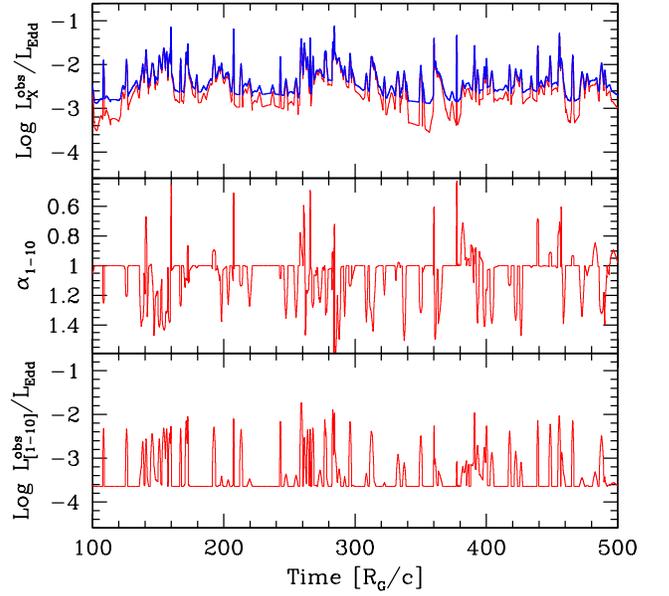,width=.53\textwidth}  
\vskip -0.7 true cm  
\caption{
Light curves of the integrated X--ray luminosity
(in units of the Eddington one)
corresponding the top panel of Fig. \ref{lc_alt},
the light curve of the spectral index and flux
between 1 and 10 keV.
We have assumed that the spectral shape of the steady
component is $F(\nu)\propto\nu^{-1}e^{-\nu/\nu_c}$.
}
\label{spectrum1}
\end{figure}

\begin{figure}  
\vskip -0.7 true cm
\psfig{file=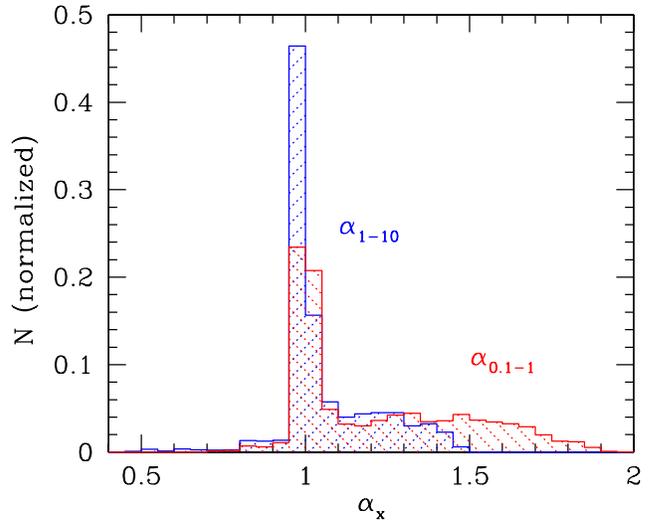,width=.53\textwidth}  
\vskip -2 true cm  
\caption{
Histograms of the soft (0.1--1 keV) and hard (1-10 keV)
X--ray spectral indices for the case shown in Fig. \ref{spectrum1}.
}
\label{isto1}
\end{figure}

Fig. \ref{spectra_r05} shows some example of spectra born out
from our simulations. 
They correspond to the case shown in the top panel
of Fig. \ref{lc_alt}, assuming a black hole of 
$10^8$ solar masses.
Spectra are calculated every 40 seconds, but are shown
every 800 seconds (corresponding to $0.53 R_{\rm g}/c$), for clarity,
from $t/(R_{\rm g}/c)=240$ to $t/(R_{\rm g}/c)=300$.
Fig. \ref{spectrum1} shows again the light curve corresponding
to the top panel of Fig. \ref{lc_alt}, together with the associated
1--10 keV spectral index, $\alpha_{[1-10]}$ 
and the light curve of the 1--10 keV flux.
Fig. \ref{isto1} shows the histograms of $\alpha_{[1-10]}$ and
the softer spectral index $\alpha_{[0.1-1]}$ (between 0.1 and 1 keV).


By construction, a spectral index different from unity
corresponds to the jet emission dominating over the steady
component.
Most of the time, the observed slopes corresponding to the jet emission are 
steeper than unity (as shown in Fig. \ref{isto1}), and even more
so at lower frequencies.

\subsection{Trends}

In this section we summarize the results of the survey of 
the parameter space aimed to test the sensitivity
of our results to our assumptions.
We indicate the values of the parameters we have changed,
all other parameters have values equal to the 
case shown in the top panel of Fig. \ref{lc_alt}.

\begin{itemize}

\item {\bf Initial velocity ---}
Fig. \ref{lc_b0} shows the effect of changing the average
initial velocity of the ejected shell
($\langle\beta_0\rangle=$0.3, 0.5 and 0.7 
from top to bottom).
It can be seen that an increase of the average initial velocity
has the effect to increase the density of spikes and their
average luminosity.
Increasing the average velocity, in fact, increase the average
number of collisions per unit time and the average efficiency
of each collision. The overall process becomes more efficient.

\item {\bf Launching site ---}
Increasing the launching site from the black hole
(we tried $R_0=3$, 5 and 7 $R_{\rm g}$)
the bolometric and the 1--10 keV light curves becomes more spiky,
with the jet dominant (in the 1--10 keV band) an increasing number 
of times over the steady component.
The net effect is, basically, the same as increasing 
the initial velocity.

\item {\bf Shell size ---}
Increasing $\Delta r_{\rm j}$ (we tried $\Delta r_{\rm j}=1$, 
2 and 4 $R_{\rm g}$, always with $r_j=4R_{\rm g}$), the light
curve becomes slightly less spiky, as the duration of the 
collisions is proportional to $\Delta r_{\rm j}$.

\item {\bf Spin of the black hole ---}
We have found negligible differences between a maximally rotating
Kerr and a Schwarzchild black hole for the launching site
$R_0=5 R_{\rm g}$.
There is still a negligible difference for the case
$R_0 =3 R_{\rm g}$ and $\beta_0=0.7$.

\item {\bf Ratio of jet to disk power ---}
We consider this as the most important parameter.
As the jet power increases, the jet emission becomes
(obviously) more and more dominant over the steady emission,
and the average spectral index flattens, to become
flatter than unity for $L_{\rm j} \sim 20 L_{\rm d}$.
For larger values of the jet power, the radiation produced
by the jet is larger that the luminosity radiated by the disk.
The radiation coming from the jet and reprocessed by the
disk can no longer be neglected, and our treatment
becomes invalid.
However, in this case, the power in the optical--UV
and in the X--ray bands are comparable, in contrast
with the majority (albeit not all) of Seyfert galaxies.

\end{itemize}

\section{Summary and discussion}  
  
We have explored the possibility that all Active Galactic Nuclei  
form jets or outflows, with a range of velocities, but with a power  
which is comparable to the power extracted by the accretion process.  
We have then assumed that in most cases (corresponding to radio--quiet sources) 
the jets are launched with velocities smaller than the escape velocity.  
Mainly for simplicity, but also in analogy with the ``internal shock"  
scenario proposed to work in radio--loud sources and in gamma--ray bursts,  
we have further assumed that the central engine works intermittently,  
producing shells or blobs.   
A shell with $\beta<\beta_{\rm esc}$ will reach a maximum distance from   
the hole, then stop and invert its motion, and may eventually collide with   
the successive shell.  
In this case the bulk kinetic energy of the two shells is   
dissipated, and the fraction of it which is given to electrons   
can be transformed into radiation.  
The accelerated electrons are embedded in the dense radiation  
field produced by the accretion disk: they cool rapidly by the   
inverse Compton process, producing the X--ray continuum.  
Simple energy balance is sufficient to estimate the Comptonization  
parameter as a function of the power dissipated by the   
colliding blobs and the disk luminosity.  
By construction, the shells are moving with velocities  
smaller than the escape speed, yet they carry a power which is  
comparable to that extracted by accretion.  
This implies that the shells are ``heavy".  
They cannot be formed by electron--positron pairs only, because   
in this case the corresponding optical depth is so large that most of  
them annihilate in less than a dynamical time.  
The required density in protons and the accompanying electrons is  
large enough to limit the importance of pairs not only  
as energy carriers, but also as scatterers.  
 
The main aim of this paper is to investigate the general properties 
of the proposed idea, to check if it can work at least at the first  
order of approximation. 
  
Our ``aborted jet" scenario is not necessarily alternative to the popular  
``disk--corona" model.  
Both processes could be active and contribute to the formation  
of the high energy continuum in the same source.
%
%
On the other hand we would like to stress that in our proposed scenario  
the source of energy could be the spin of the hole, besides accretion.  
Pushing this possibility to the limit (i.e. all the high energy emission 
produced by AGNs comes from the rotational energy of their black hole),  
would result in the remarkable fact that it is the black hole spin,   
rather than accretion, which produces the bulk of the X--ray background.  
It is then instructive to isolate the ``aborted jet" process  
in order to find ways to confirm or falsify this scenario.  
  
One of the clearest difference with the disk--corona model   
is that the dissipation of energy should occur along the  
axis of rotation of the black hole.   
This implies that the X--ray flux coming from the colliding shells  
will illuminate preferentially the inner part of the disk, especially  
when they collide close to the hole.  
This may solve the problem of the formation of the strong red wings   
of the relativistic iron line observed in MCG--6--30--15 (Wilms et al. 2001),  
which requires an ``illuminator" emissivity strongly increasing  
towards the black hole. It should be noted that we have neglected, for 
simplicity,  the light bending due to the strong gravity   
(see e.g. Martocchia et al. 2002), which results in an enhanced illumination  
of the innermost disk regions. 
The illumination could be further enhanced by anisotropic Compton scattering  
(since the seed photons are coming from the disk, more inverse Compton  
radiation is channeled back towards the disk than along the viewing angle, see  
e.g. Ghisellini et al. 1991; Malzac et al. 1998). 
Another cause of anisotropy is beaming of the X--ray radiation,
which is preferentially emitted towards the accretion disk in 
efficient shell--shell collisions.
According to our simulations, in fact, the most efficient
collisions are between massive blobs coming back to the disk and having
already experienced some collisions, and newly generated 
blobs moving in the opposite direction.
While this may help explaining the large equivalent widths of Fe lines 
observed in a few cases (notably MCG--6--30-15), it is apparently at odds 
with the relative paucity of relativistic iron lines observed by XMM--Newton 
(e.g. Reeves et al. 2003).   
However, it should be noted that the increase of illuminating X--ray   
photons may result in a significant ionization of the innermost regions   
of the accretion disc, making predictions on the iron line intensity less   
straightforward (e.g. Nayakshin \& Kazanas 2002, and references therein).   
Detailed calculations of the iron line properties  
are beyond the scope of this paper, and are deferred to a future work.  
 
We note that an important piece of information may come from the observations 
of the Compton reflection continuum and iron line in radio--loud sources. 
If the jet in these sources is successful, in fact, it should not illuminate 
much of the accretion disk, and therefore these objects should have 
weaker reflection features produced by the corona only. 
This seems indeed to be the case
(e.g. Grandi et al. 2002, and references therein). 
Then the equivalent width of the fluorescent iron lines in radio--galaxies  
may measure the importance of the corona with respect to the jet 
in producing the thermal X--ray continuum, once the data are purified 
from all other additional contributions (e.g. the non thermal  
radiation from the jet).


The spectral index of the jet emission, calculated in our simulations,
is generally steeper than the average spectral index observed 
in Seyfert galaxies (i.e. $\alpha_x\sim 1$).
As it is, our model requires therefore the presence of a steadier
component, with the ``right" spectral index, contributing
to X--ray band.
This steady component should have a bolometric luminosity
which is, on average, smaller than the average power of the jet,
even if its relative contribution in the 1--10 keV band 
is more important.
The steep jet emission, when contributing notably
to the 1--10 keV band, would steepen the overall spectral index
and increase the flux.
It would then produce a ``steeper when brighter" behavior
as observed in Seyfert galaxies (e.g. Zdziarski et al. 2003).
Occasionaly, instead, the jet emission is both dominant and
characterized by a flat spectrum, and we have then the opposite 
behavior, i.e. "harder when brighter", but this occurs
more rarely.

As it is, our model explains the X--ray properties of
Narrow Line Seyfert 1 galaxies (Boller et al. 1996; Brandt et al. 1997; 
Cancelliere \& Comastri 2002). 
These sources are in fact characterized, on one hand,
by a 2--10 keV spectral index between 1 and 1.5 (and an even steeper spectrum
in the softer band), and, on the other hand, by  a short term, 
large amplitude variability.
It is then possible than the main difference between Narrow Line Sey1 
(including in this class also sources like MCG--6--30--15 which have 
broad lines but in X--rays behaves like NLSy1s)
and classical Seyferts is the ratio between jet and disc/corona emission. 
It is worth noting that NLSy1 are widely believed to have a larger 
$L/L_{\rm Edd}$ ratio than classical Seyferts, which again can be explained 
by an enhanced jet emission. 
If this is true, one could speculate that the physical parameter behind 
the NLSy1 X--ray behaviour is not (or at least not only) the accretion rate, as 
usually supposed, but the presence of a more powerful aborted jet.


Regarding broad line, classical Seyfert 1 galaxies,
we should however consider that our model neglects,
for simplicity and ease of calculation, a few important
physical effects.
One of these concerns light bending, important when
the emitting spot is very close to the black hole.  
This effect is expected to change the  
observed X--ray luminosities only by a factor of a few 
(Martocchia et al. 2002), 
but that can nevertheless be very important for a detailed 
study of spectral evolution. 
Indeed, a different degree of light bending and gravitational 
redshift corresponding to different heights of the illuminator
above the black hole can explain the puzzling temporal 
behavior of MCG--6--30--15, where the continuum and iron 
line variabilities are decoupled (Miniutti et al. 2003).

Another important effect, neglected here,
is the feedback between the luminosity
produced by the jet and the disk emission,
important for large ratios between the jet and the disk powers.
In these cases the radiation reprocessed by the disk can
become important and introduce the same kind of feedback 
which makes the hot corona model to work, producing 
spectral indices close to unity in the X--ray band.

Finally, we would like to comment about the difference   
between radio--loud and radio--quiet sources.  
In our scenario, this is mainly a difference in mass loading,  
coupled with a possible difference in jet power.  
The central engine in radio--loud sources succeeds in accelerating  
jets at speeds larger than the escape velocity: in these sources  
the jet power can dominate the total energetics (as in BL Lac objects),  
and the outflow mass rate is of the order of a per cent of the   
accretion rate.  
The ``jet" of radio--quiet sources may not be much  
less powerful than in radio--loud objects, if it  
contributes significantly to the formation of the X--ray flux.  
What should be different is the outflowing mass rate, 
which must be greater in radio--quiet objects, making their ``jets" 
move slower. 
If all jets are powered by the extraction of rotational 
energy from a spinning black hole, it is then possible that  
it is this mechanism, and not accretion, to be responsible for 
all the high energy radiation produced by AGNs 
(i.e. all the X--ray and the $\gamma$--ray flux). 
 
It is also possible that a specific source, usually radio--quiet, 
occasionally may launch ``successful" shells, with relativistic speeds.  
However, these ``successful jet episodes" 
must be rare in AGN, since we rarely see ``fossil" 
long lived weak radio lobes in not jetted sources. 
This may occur more often in galactic micro--quasars,  
and be associated with the major radio--flares. 
The bulk Lorentz factor associated with 
major radio events in GRS 1915+105 is relatively small, 
perhaps suggesting that, when radio--weak, the jet 
is not successfully launched because it does not attain 
bulk speeds larger than the escape velocity. 
These sources may therefore be the ``missing link"  
between radio--loud and radio--weak sources, changing 
from time to time from one class to the other. 
 

\begin{acknowledgements}  
We thank M. Abramowicz, V. Karas and D. Malesani for useful discussions
and the anonymous referee for having been at the same time severe 
and encouraging.
We thank MIUR and ASI for funding. 
\end{acknowledgements}

\end{document}